\documentclass[a4paper,12pt]{article}

\usepackage{graphics,psfrag}
\usepackage{amsmath,amsthm,amssymb,epsfig,euscript,array,cite,cancel,color}
\usepackage{mathrsfs}
\usepackage{cases,empheq}
\usepackage[colorlinks=false]{hyperref}
\usepackage{verbatim}
\usepackage{ulem}
   \usepackage[vcentermath]{youngtab}   % Young tableaux
   
   \usepackage[usenames,dvipsnames]{xcolor}  % additional colors 
\usepackage[font=small]{caption} % change caption size

\usepackage[paper=a4paper,dvips,top=2.54cm,left=2.74cm,right=2.34cm,
    foot=1cm,bottom=2.54cm]{geometry}

\theoremstyle{definition} % %

\theoremstyle{definition}
\numberwithin{trial}{subsection}

\numberwithin{dtl}{subsection}

\theoremstyle{remark}

\newcounter{multieqs}

\newcommand{\be}{\begin{equation}}
\newcommand{\ee}{\end{equation}}
\newcommand{\eq}[1]{(\ref{#1})}
\newcommand{\bit}{\begin{itemize}}  \newcommand{\eit}{\end{itemize}}

\newcommand{\bm}[1]{\mbox{\boldmath $#1$}}
\newcommand{\rf}[1]{(\ref{#1})}

\def\bd{\begin{document}}
\def\ed{\end{document}}
\def\nn{\nonumber}
\def\bea{\begin{eqnarray}}
\def\eea{\end{eqnarray}}
\let\bm=\bibitem

\def\la{\langle}
\def\ra{\rangle}

\def\npb#1#2#3{Nucl. Phys. {\bf{B#1}} #3 (#2)}
\def\plb#1#2#3{Phys. Lett. {\bf{#1B}} #3 (#2)}
\def\prl#1#2#3{Phys. Rev. Lett. {\bf{#1}} #3 (#2)}
\def\prd#1#2#3{Phys. Rev. {D \bf{#1}} #3 (#2)}
\def\cmp#1#2#3{Comm. Math. Phys. {\bf{#1}} #3 (#2)}
\def\cqg#1#2#3{Class. Quantum Grav. {\bf{#1}} #3 (#2)}
\def\nppsa#1#2#3{Nucl. Phys. B (Proc. Suppl.) {\bf{#1A}}#3 (#2)}
\def\ap#1#2#3{Ann. of Phys. {\bf{#1}} #3 (#2)}
\def\ijmp#1#2#3{Int. J. Mod. Phys. {\bf{A#1}} #3 (#2)}
\def\rmp#1#2#3{Rev. Mod. Phys. {\bf{#1}} #3 (#2)}
\def\mpla#1#2#3{Mod. Phys. Lett. {\bf A#1} #3 (#2)}
\def\jhep#1#2#3{J. High Energy Phys. {\bf #1} #3 (#2)}
\def\atmp#1#2#3{Adv. Theor. Math. Phys. {\bf #1} #3 (#2)}

\def\N{{\cal N}}
\def\sst{\scriptscriptstyle}
\def\thetabar{\bar\theta}
\def\Tr{{\rm Tr}}
\def\one{\mbox{1 \kern-.59em {\rm l}}}

\def\a{\alpha}      \def\da{{\dot\alpha}}  \def\dA{{\dot A}}
\def\b{\beta}       \def\db{{\dot\beta}}  
\def\g{\gamma}  \def\G{\Gamma}  \def\dc{{\dot\gamma}}  
\def\d{\delta}  \def\D{\Delta}  \def\ddt{\dot\delta}  
\def\e{\epsilon}        \def\ve{\varepsilon}  
\def\f{\phi}    \def\F{\Phi}    \def\vvf{\f}  
\def\h{\eta}  
\def\k{\kappa}  
\def\l{{\lambda}} \def\L{\Lambda}  
\def\m{\mu} \def\n{\nu}  
\def\om{\omega}  \def\Om{\Omega}  
\def\p{\pi} \def\P{\Pi}  
\def\r{\rho}  
\def\s{\sigma}  \def\S{\Sigma}  
\def\t{\tau}  
\def\th{\theta} \def\Th{\Theta} \def\vth{\vartheta}  
\def\X{\Xeta}  
\def\z{\zeta}  

\def\na{\nabla}  
\def\cA{{\cal A}} \def\cB{{\cal B}} \def\cC{{\cal C}}  
\def\cD{{\cal D}} \def\cE{{\cal E}} \def\cF{{\cal F}}  
\def\cG{{\cal G}} \def\cH{{\cal H}} \def\cI{{\cal I}}  
\def\cJ{{\cal J}} \def\cK{{\cal K}} \def\cL{{\cal L}}  
\def\cM{{\cal M}} \def\cN{{\cal N}} \def\cO{{\cal O}}  
\def\cP{{\cal P}} \def\cQ{{\cal Q}} \def\cR{{\cal R}}  
\def\cS{{\cal S}} \def\cT{{\cal T}} \def\cU{{\cal U}}  
\def\cV{{\cal V}} \def\cW{{\cal W}} \def\cX{{\cal X}}  
\def\cY{{\cal Y}} \def\cZ{{\cal Z}}

\def\ua{{\underline{\alpha}}} 
 \def\ub{\underline{\phantom{\alpha}}\!\!\!\beta}  
\def\uc{\underline{\phantom{\alpha}}\!\!\!\gamma}  
\def\um {{\underline{\mu}}}  
\def\ud{{\underline{\delta}}} 
\def\ue{\underline\epsilon}  
\def\una{{\underline a}}\def\uA{{\underline A}}  
\def\unb{{\underline b}}\def\uB{{\underline B}} 
\def\unc{{\underline c}}\def\uC{{\underline C}}  
\def\und{{\underline d}}\def\uD{{\underline D}}  
\def\une{{\underline e}}\def\uE{{\underline E}}  
\def\unf{{\underline{\phantom{e}}\!\!\!\! f}}\def\uF{{\underline F}}  
\def\unm{{\underline m}\def\uM{\underline M}} 
\def\unn{{\underline n}\def\uN{\underline N}} 
\def\unp{{\underline{\phantom{a}}\!\!\! p}}\def\uP{{\underline P}}  
\def\unq{{\underline{\phantom{a}}\!\!\! q}}  
\def\uQ{{\underline{\phantom{A}}\!\!\!\! Q}}  
\def\uH{{\underline{H}}}  
\def\uM{{\underline{M}}}
\def\uN{{\underline{N}}}
  
\def\As {{A \hspace{-6.4pt} \slash}\;}  
\def\bs {{b \hspace{-6.4pt} \slash}\;}  
\def\Ds {{D \hspace{-6.4pt} \slash}\;}
\def\Gts {{\Gt \hspace{-6.4pt} \slash}\;}
\def\ds {{\del \hspace{-6.4pt} \slash}\;}  
\def\ss {{\s \hspace{-6.4pt} \slash}\;}  
\def\ks {{ k \hspace{-6.4pt} \slash}\;}  
\def\ps {{p \hspace{-6.4pt} \slash}\;}   
\def\xs {{x \hspace{-6.4pt} \slash}\;}  
\def\pas {{{p_1} \hspace{-6.4pt} \slash}\;}  
\def\pbs {{{p_2} \hspace{-6.4pt} \slash}\;}   
\def\cFs {{{\cal F} \hspace{-6.4pt} \slash}\;}

\def\Ah{{\hat{A}}}  
\def\Dh{{\hat{D}}}
\def\Gh{{\hat{G}}}
\def\Fh{{\hat{F}}}
\def\Ih{{\hat{I}}} 
\def\Jh{{\hat{J}}} 
\def\Kh{{\hat{K}}}
\def\Lh{{\hat{L}}} 
\def\Ph{{\hat{P}}}
\def\Rh{{\hat{R}}}
\def\Vh{{\hat{V}}} 
\def\Xh{{\hat{X}}}
 
\def\ah{{\hat{a}}}
\def\bh{{\hat{b}}}
\def\ch{{\hat{c}}}
\def\gh{{\hat{g}}}
\def\dh{{\hat{d}}}
\def\hh{{\hat{h}}}
\def\uh{{\hat{u}}}  
\def\vh{{\hat{v}}}
\def\xh{{\hat{x}}} 
\def\yh{{\hat{y}}}
\def\zh{{\hat{z}}}
\def\ph{{\hat{p}}}
\def\qh{{\hat{q}}}
\def\thh{{\hat{t}}}  
\def\xih{\hat{\xi}}  
\def\Psih{\hat{\Psi}}    
\def\mh{{\hat{m}}}
\def\nh{{\hat{n}}}
\def\ih{{\hat{i}}}
\def\jh{{\hat{j}}}
\def\kh{{\hat{k}}}
\def\aah{{\hat{\alpha}}}
\def\bbh{{\hat{\beta}}}
\def\ggh{{\hat{\gamma}}}
\def\llh{{\hat{\ell}}} 
\def\ph{{\hat{p}}}

\def\psit{\tilde{\psi}}  
\def\Psit{\tilde{\Psi}}   
\def\Psibt{\tilde{\bar{Psi}}}  
\def\st{\tilde{\sigma}}  
\def\delt{\tilde{\delta}}
\def\Phit{\tilde{\Phi}}   
\def\Phitb{\overline{\tilde{Phi}}}  
\def\tht{\tilde{\th}}  
\def\lt{\tilde{\l}}
\def\chit{\tilde{\chi}}   
\def\phit{\tilde{\phi}} 
\def\omt{\tilde{\omega}}
\def\Gmt{\tilde{\Gamma}}

\def\At{\tilde{A}}
\def\Bt{\tilde{B}}
\def\Ct{\tilde{C}}
\def\Dt{\tilde{D}}
\def\Et{\tilde{E}}
\def\Ft{\tilde{F}}
\def\Gt{\tilde{G}}
\def\Ht{\tilde{H}}
\def\It{\tilde{I}}
\def\Jt{\tilde{J}}
\def\Qt{\tilde{Q}}  
\def\Rt{\tilde{R}}  
\def\Mt{\tilde{M }}  
\def\Nt{\tilde{N}}   
\def\St{\tilde{S}}
\def\Vt{\tilde{V}}
\def\Xt{\tilde{X}} 
\def\at{\tilde{a}}
\def\ct{\tilde{c}}
\def\dt{\tilde{d}}
\def\htt{\tilde{h}} 
\def\ft{\tilde{f}}
\def\gt{\tilde{g}}
\def\pt{\tilde{p}}  
\def\qt{\tilde{q}}  
\def\vt{\tilde{v}}  
\def\nt{\tilde{n}}  
\def\ut{\tilde{u}}  
\def\wt{\tilde{w}}  
\def\zt{\tilde{z}} 
\def\xt{\tilde{x}} 
\def\yt{\tilde{y}} 
\def\Psit{\tilde{\Psi}}
\def\vphit{\tilde{\varphi}}

\def\gamt{\tilde{\gamma}}

\def\Tt{\tilde{T}}

\def\ot{{\tilde{\omega}}}
\def\eb{\bar{\epsilon}} 
\def\delb{\bar{\partial}}  
\def\thb{\bar{\theta}}
\def\Thb{{\bar{\Theta}}}
\def\mub{\bar{\mu}}
\def\lamb{\bar{\l}}
\def\psib{\bar{\psi}}
\def\sb{\bar{\sigma}}
\def\xib{\bar{\xi}}
\def\chib{\bar{\chi}}

\def\Psib{\bar{\Psi}}
\def\Phib{\bar{\Phi}}
\def\Lamb{\bar{\Lambda}}
\def\Sb{{\overline \Sigma}}
\def\cb{\bar{c}}
\def\hb{\bar{h}}
\def\qb{\bar{q}}
\def\wb{\bar{w}}
\def\zb{{\bar{z}}}
\def\Hb{\bar{H}}
\def\Qb{{\bar Q}}
\def\Omegab{\overline{\Omega}}
\def\ob{\overline{\omega}}
\def\Gab{{\bar{\Gamma}}}

\def\Ab{{\overline A}} \def\Bb{{\overline B}} \def\Cb{{\overline C}}  
\def\Db{{\overline D}} \def\Eb{{\overline E}} \def\Fb{{\overline F}}  
\def\Gb{{\overline G}} 
\def\Ib{{\overline I}}  
\def\Jb{{\overline J}} \def\Kb{{\overline K}} \def\Lb{{\overline L}}  
\def\Mb{{\overline M}} \def\Nb{{\overline N}} \def\Ob{{\overline O}}  
\def\Pb{{\overline P}}  \def\Rb{{\overline R}}  
 \def\Tb{{\overline T}} \def\Ub{{\overline U}}  
\def\Vb{{\overline V}} \def\Wb{{\overline W}} \def\Xb{{\overline X}}  
\def\Yb{{\overline Y}} \def\Zb{{\overline Z}}  

\def\fb{{\overline f}}
\def\gb{{\overline g}}
\def\mb{{\overline m}}
\def\lb{{\overline l}}
\def\yb{{\overline y}}
  
\def\ldel{{\overleftarrow{\del}}}
\def\rdel{{\overrightarrow{\del}}}
\def\ldeldel{{\overleftarrow{\del^2}}}
\def\rdeldel{{\overrightarrow{\del^2}}}
\def\ldelb{{\overleftarrow{\bar{\del}}}}
\def\rdelb{{\overrightarrow{\bar{\del}}}}
\def\ba{{\bf a}} 
\def\bk{{\bf k}}  
\def\bl{{\bf l}}  
\def\bp{{\bf p}}  
\def\bq{{\bf q}}  
\def\br{{\bf r}}
\def\bt{{\bf t}}
\def\bu{{\bf u}}
\def\bv{{\bf v}}
\def\bx{{\bf x}}  
\def\by{{\bf y}}  
\def\bR{{\bf R}}  
\def\bV{{\bf V}}  
\def\bK{{\bf K}}

\def\bone{{\bf 1}}  

\def\va{{\vec a}}
\def\vk{{\vec k}}
\def\vp{{\vec p}}
\def\vq{{\vec q}}
\def\vx{{\vec x}}
\def\vy{{\vec y}}
\def\vu{{\vec u}}
\def\vv{{\vec v}}

\def\vs{{\vec \sigma}}
\def\vtau{{\vec \tau}}

\newcommand{\ov}[1]{\overrightarrow{#1}}

\def\frA{\mathfrak{A}}
\def\frB{\mathfrak{B}}
\def\frC{\mathfrak{C}}
\def\frD{\mathfrak{D}}
\def\frE{\mathfrak{E}}
\def\frF{\mathfrak{F}}
\def\frG{\mathfrak{G}}
\def\frH{\mathfrak{H}}
\def\frM{\mathfrak{M}}
\def\frN{\mathfrak{N}}
\def\frR{\mathfrak{R}}
\def\frW{\mathfrak{W}}

\def\fra{\mathfrak{a}}
\def\frb{\mathfrak{b}}
\def\frf{\mathfrak{f}}
\def\frg{\mathfrak{g}}
\def\frh{\mathfrak{h}}
\def\frl{\mathfrak{l}}
\def\frs{\mathfrak{s}}
\def\fri{\mathfrak{i}}
\def\frj{\mathfrak{j}}

\def\ma{\mathfrak{a}}
\def\mg{\mathfrak{g}}
\def\mR{\mathfrak{R}}
\def\mN{\mathfrak{N}}

\def\d{\delta}\def\D{\Delta}\def\ddt{\dot\delta}  
  
\def\pa{\partial} \def\del{\partial}  
\def\xx{\times}  
\def\uno{\mbox{1 \kern-.59em {\rm l}}}    
  
\def\trp{^{\top}}  
\def\inv{^{-1}}  
\def\dag{{^{\dagger}}}  
\def\pr{^{\prime}}  
  
\def\rar{\rightarrow}  
\def\lar{\leftarrow}  
\def\lrar{\leftrightarrow}  
  
\newcommand{\0}{\,\!}      %this is just NOTHING!  
\def\one{1\!\!1\,\,}  
\def\im{\imath}  
\def\jm{\jmath}  
  
\newcommand{\tr}{\mbox{tr}}  
\newcommand{\slsh}[1]{/ \!\!\!\! #1}  
  
\def\vac{|0\rangle}  
\def\lvac{\langle 0|}  
  
\def\hlf{\frac{1}{2}}  
\def\ove#1{\frac{1}{#1}}  

\def\Box{\square}  
\def\CC {\mathbb{C}}
\def\FF {\mathbb{F}}
\def\RR{\mathbb{R}}
\def\NN{\mathbb{N}}  
\def\ZZ{\mathbb{Z}}  
\def\bb#1{{\bf #1}}  
\def\bcomment#1{}  
\def\bfhat#1{{\bf \hat{#1}}}  
\def\VEV#1{\left\langle #1\right\rangle}  

\newcommand{\ex}[1]{{\rm e}^{#1}} \def\ii{{\rm i}}  

\newcommand{\lrbrk}[1]{\left(#1\right)}
\newcommand{\lrsbrk}[1]{\left[#1\right]}
\newcommand{\lrcbrk}[1]{\left\{#1\right\}}
\newcommand{\sfrac}[2]{{\textstyle\frac{#1}{#2}}}
 
\def\stw{{\sqrt{2}}}

\def\rf {{\rm f}}
\def\ri {{\rm i}}
\def\rj {{\rm j}}
\def\rk {{\rm k}}
\def\rl {{\rm l}}
\def\rs {{\scriptscriptstyle \rm S}}
\def\rt {{\scriptscriptstyle \rm T}}

\def\rQ {{\scriptscriptstyle \rm \cQ}}
\def\rR {{\scriptscriptstyle \rm \cR}}

\def\cQb{{\cal \Qb}}
\def\cRb{{\cal \Rb}}
\def\cWb{{\cal \Wb}}

\def\fd {{\rm N}}
\def\afd {{\overline{\rm N}}}

\def \II {I\hspace{-.1em}I\hspace{.1em}}
\def \IIA {\mbox{\II A\hspace{.2em}}}
\def \IIB {\mbox{\II B\hspace{.2em}}}
\def \gs {g^s}
\def \ls {\lambda^s}

\def \I {{\cal I}}
\def \qs {q\hspace{-.53em}/\hspace{.15em}}
\def \ks {k\hspace{-.53em}/\hspace{.15em}}
\def \YM {{\mbox{\tiny YM}}}
\def \gym {g_{\YM}}

\def \Lc {\L_c}
\def\IR{\relax{\rm I\kern-.18em R}}
\def \id {{\bf 1}}

\def\cci{\ell}
\def\ccj{\ell'}

\def \thbb{\overline{\th\th}}
\newcommand \ol{\overline}
\def \lamb{\bar{\lambda}}
\def \vphi{\varphi}
\def \lambh{\hat{\bar{\lambda}}}
\def \lh{\hat{\lambda}}
\def \dd{\ddagger}

\def \ad {\dot{a}}
\def \bd {\dot{b}}
\def \cd {\dot{c}}
\def  \ddd {\dot{d}}
\def \ed {\dot{e}}
\def \fd {\dot{f}}
\def \Bh {\hat{B}}
\def \zm {{(0)}}
\def \nz {{(\text{KK})}}
\def \3{{(3)}}
\def \diag {\text{diag}}
\def \inm {{(m^{-1})}}
\def \3{{(3)}} 
\def \6{{(6)}}
\def \2{{(2)}}
\def \7{{(7)}} 
\def \4{{(4)}}
\def\1{{(1)}}
\def\5{{(5)}}
\def\0{{(0)}}

\def\eh{{\hat{e}}}
\def\fh{{\hat{f}}}
\def\lh{{\hat{l}}}
\def\rh{{\hat{r}}}
\def\wh{{\hat{w}}}
\renewcommand{\mh}{{\hat{m}}}

\def \DBI{{\text{DBI}}}
\def\et{{\tilde{\e}}}
\def\w{{\wedge}}
\def\bbV{{\mathbb{V}}}
\def\M{{(\text{M})}}
\def\T{{(\text{T})}}

\def\Hbt{{\tilde{\bar{H}}}}
\def\Fbt{{\tilde{\bar{F}}}}

\def\fR{{\mathfrak{R}}}
\def\fg{{\mathfrak{g}}}
\def \sk {\textsc{k}}
\def\S{{\Sigma}}
\def\nb{{\nabla}}
\def\bB{{\bf B}}

\colorlet{1}{red}
\colorlet{2}{green}
\colorlet{3}{blue}
\colorlet{4}{cyan}
\def\bJ{{\bold J}}
\def\tR{{\text{(R)}}}
\def\bQ{{\bold Q}}
\def\bo{{\pmb{\omega}}}
\def\ADM{{\text{ADM}}}
\def\st{{\tilde\sigma}}
\def\Fbh{{\Phi_{\text{bh}}}}
\def\bC{{\bold C}}
\def\bep{{\pmb\epsilon}}
\def\EM{{(\text{EM})}}
\def \bTh {\bold \Theta}
\def\bL{{\bold L}}
\def\tG{{\tilde{\Gamma}}}
\def \vL {\mathfrak{L}}
\def\bE{{\bold E}}
\def\nbs{{\overset{*}{\nb}}}
\def\nbt{{\tilde{\nb}}}

\def\pl{\partial}
\def\tt{\th_{\mbox{\tiny T}}}
\def\tl{\th_{\mbox{\tiny L}}}
\def\Lef{\Lambda_{\mbox{\tiny eff}}}
\def\lef{\ell_{\mbox{\tiny eff}}}
\def\OmH{\Omega_{\mbox{\tiny H}}}
\def\TH{T_{\mbox{\tiny H}} }
\def\kH{\kappa_{\mbox{\tiny H}} }
\def\in{{\mbox i}_\xi}
\def\iz{{\mbox i}_\zeta}
\def\dr{{\mbox d}}
\def\Dr{{\mbox D}}
\def\nd{{\nabla}}
\def\ndt{{\tilde \nabla}}

\def \rd {\color[rgb]{1,0,0}}
\def \gr {\color[rgb]{0,0.55,0}}
\def \bl {\color[rgb]{0,0.1,1}}
\def \pr {\color[rgb]{0.9,0,0.9}}

\author{Baoyi Chen\footnote{baoyi@tapir.caltech.edu}$~^a$, Feng-Li Lin\footnote{linfengli@phy.ntnu.edu.tw}$~^b$
and Bo Ning\footnote{ningbo@scu.edu.cn}$~^c$ 
\\
\\
{\small\it $^a$ TAPIR, Walter Burke Institute for Theoretical Physics,} 
\\
{\small\it  California Institute of Technology, Pasadena, California 91125, USA}
\\
{\small\it $^b$ Department of Physics, National Taiwan Normal University,}
\\
{\small\it No. 88, Sec. 4, Ting-Chou Road, Taipei 11677, Taiwan}
\\
{\small\it $^c$  College of Physical Science and Technology, }
\\
{\small\it Sichuan University, Chengdu, Sichuan 610065, China}
}

\title{\bf {\Large Gedanken Experiments to Destroy a BTZ Black Hole} }

\date{}
\begin{document}

\maketitle

\thispagestyle{empty}
\newpage

\abstract{ 

\hspace{2mm}

We consider gedanken experiments to destroy an extremal or near-extremal BTZ black hole by throwing matter into the horizon.  These black holes are vacuum solutions to (2+1)-dimensional gravity theories, and are asymptotically $\mathrm{AdS}_3$.
Provided the null energy condition for the falling matter, we prove the following---(i) in a Mielke-Baekler model without ghost fields, when torsion is present, an extremal BTZ black hole can be overspun and becomes a naked conical singularity; 
(ii) in 3-dimensional Einstein gravity and chiral gravity, which both live in the torsionless limits of Mielke-Baekler model, an extremal BTZ black hole cannot be overspun; 
and (iii) in both Einstein gravity and chiral gravity, a near-extremal BTZ black hole cannot be overspun, leaving the weak cosmic censorship preserved.   
To obtain these results, we follow the analysis of Sorce and Wald on their gedanken experiments to destroy a Kerr-Newman black hole, and calculate the second order variation of   the black hole mechanics.   
Furthermore,  Wald's type of gedanken experiments provide an operational procedure of proving the third law of black hole dynamics. Through the AdS/CFT correspondence,  our results on BTZ black holes also indicate that a third law of thermodynamics holds for the holographic conformal field theories dual to 3-dimensional Einstein gravity and chiral gravity.

\thispagestyle{empty}
\newpage

\tableofcontents

%\setcounter{equation}0

%%%%%%%%%%%%%%%%%     1 Intro     %%%%%%%%%%%%%%%%%% %% 

 \section{Introduction} \label{sec:intro}

 Weak cosmic censorship conjecture (WCCC) was formulated by Penrose \cite{Penrose:1969pc} to postulate that a gravitational singularity should not be naked and should be hidden inside a black hole horizon. A gravitational singularity is usually mathematically ill-defined due to the divergent spacetime curvature. Thus, the WCCC helps to avoid seeing such unphysical part of the universe and retains the predicted power of physical laws. Its philosophical incarnation was summarized by Hawking that ``Nature abhors a naked singularity" \cite{Hawking:1997}.  In this sense, a special case worthy of consideration is the 3-dimensional Banados-Teitelboim-Zanelli (BTZ) black hole, for which there is no curvature singularity but a conical one. The conical singularity will thus cause no physical divergence as the curvature one.  It is then interesting to check if WCCC holds for this case or not, and partly motivates the study of this paper.

  The general proof or demonstration of WCCC is notoriously difficult. One way is to find the critical situation in which a black hole almostly turns into a naked singularity by subjecting to small perturbations. This is when a Kerr-Newman black hole is in its near-extremal regime. A super-extremal solution possesses the naked singularity, thus checking WCCC is to see if a sub-extremal black hole in the near-extremal limit can turn into a super-extremal one by throwing some matter.  Along this line of thought, a gedanken experiment was firstly proposed by Wald \cite{Wald:I} to demonstrate the impossibility of destroying an extremal Kerr-Newman black hole by throwing the matter obeying the null energy condition. The key ingredient in \cite{Wald:I} is the linear variation of black hole mechanics \cite{Toth:2011ab,Natario:2016bay}, i.e.,
\be\label{VaI}
\delta M-\Omega_H \delta J-\Phi_H \delta Q \ge 0
\ee  
where $M$ is the mass of the black hole, $J$ the angular momentum, $Q$ the charge, and $\Omega_H$ and $\Phi_H$ are respectively the angular velocity and chemical potential evaluated on the horizon. A similar consideration for the near-extremal Reissner-Nordstrom black hole was examined by Hubeny \cite{Hubeny:1998ga} and found that it can be overcharged to violate WCCC by throwing a charged particle. See \cite{deFelice:2001wj,Chirco:2010rq,Saa:2011wq,Gao:2012ca} for the follow-up works.

   Recently, it was realized by Sorce and Wald \cite{Sorce:2017dst} that the analysis in Hubeny's type of gedanken experiments is insufficient at the linear order so that the second order variation must be taken into account to check WCCC for near-extremal black holes. Based on an earlier development of  the second order variation of black hole mechanics \cite{Hollands:2012sf}, they went beyond the first order analysis in \cite{Hubeny:1998ga} and derived the following inequality
\be\label{VaII}
\delta^2 M-\Omega_H \delta^2 J -\Phi_H\delta^2 Q\ge - T_H \delta^2 S 
\ee   
with $T_H$ the Hawking temperature and $S$ the Wald's black hole entropy  \cite{Wald:1993nt}\cite{Iyer:1994ys}, which equals to the Bekenstein-Hawking entropy of area law for the case of Einstein's theory of gravity,  but receives modifications for non-Einstein theories of gravity\cite{Jacobson:1993vj} (See (\ref{entropy}) for the case of Mielke-Barkler gravity).
Under the situation that the linear variation is optimally done, i.e.,  the inequality (\ref{VaI}) is saturated, Sorce and Wald adopted (\ref{VaII}) to show that the WCCC holds for Kerr-Newman black holes in 4-dimensional Einstein-Maxwell gravity. In \cite{Sorce:2017dst} it is assumed that the near-extremal black hole is linearly stable, so that at very late time the linear perturbation induced by falling matter becomes the perturbation towards another Kerr-Newman  black hole. Thus, the 
WCCC is to prohibit the possibility of a naked singularity, and can be formally described as the condition for a 1-parameter family of black hole solutions
\be
f(\lambda)>0, \qquad \textrm{for all} \,\, \lambda\ge 0
\ee
with $f(\lambda)=0$ being the condition for extremal black hole. For examples,  $f(\lambda)=M(\lambda)^2-{J(\lambda)^2\over M(\lambda)^2}-Q(\lambda)^2$ for a Kerr-Newman black hole of mass $M(\lambda)$, angular momentum $J(\lambda)$ and charge $Q(\lambda)$, and $f(\lambda)=M(\lambda)^2+\Lambda J(\lambda)^2$ for a BTZ black hole in 3-dimensional anti-de Sitter (AdS) space of cosmological constant $\Lambda<0$. Note that there is no need in this formulation of examining WCCC to consider the self-force effects of the in-falling matter as done in \cite{Barausse:2010ka,Barausse:2011vx,Zimmerman:2012zu,Colleoni:2015afa,Colleoni:2015ena}.

    In this paper, we will check WCCC for a BTZ black hole in 3-dimensional torsional Mielke-Barkler gravity (MBG) \cite{Mielke:1991nn,Baekler:1992ab,Hehl:1976kj} with the general falling  matter \footnote{See also recent papers \cite{Duztas:2018fbc,Chen:2018yah} for the related discussion for special falling matter.}. In some special limits of MBG we have either Einstein gravity or chiral gravity \cite{Li:2008dq}, both of which have the known dual descriptions by a 2-dimensional conformal field theory (CFT) in the context of AdS/CFT correspondence \cite{Maldacena:1997re}. Especially, the extremal black hole has zero surface gravity, and corresponds to a dual CFT state at zero temperature. The motivation of our study is two folds. First, we would like to see if WCCC holds even for the naked conical singularity such as the one in BTZ, and at the same time extend the formulation of \cite{Sorce:2017dst} to more general gravity theories. Second, Wald's type of gedanken experiments provide an operational procedure of proving the third law of black hole dynamics \cite{Israel:1986gqz,Chirco:2010rq}: One cannot turn the non-extremal black hole into an extremal one in the finite time-interval by throwing into the black hole the matter satisfying the null energy condition. We can turn the above third law into the one of black hole thermodynamics if we adopt Bekenstein and Hawking's point of view. Moreover, through the AdS/CFT correspondence, this third law also corresponds to the third law of thermodynamics for the dual 2-dimensional CFT \footnote{See \cite{DHoker:2009ixq} for the earlier discussion for AdS$_5$ case in the context other than WCCC.}. Our results indicate that such a third law of thermodynamics holds for the holographic CFTs dual to 3-dimensional Einstein gravity and chiral gravity. Intuitively, the cooling procedure can be holographically understood as throwing the coolant, i.e., matter of spin $J$ and energy $E$ with $J>E$, into the black hole. 
    
    We organize the rest of the paper as follows. In Sec.~\ref{sec:2} we derive the linear and second order variational identities for the MB model, with which we can proceed  to the consideration of gedanken experiments for three ghost-free limits of MB model, i.e., the Einstein gravity, chiral gravity and torsional chiral gravity. In Sec.~\ref{sec:3} we consider the gedanken experiments for an extremal BTZ black hole by using the linear variational identity and the null energy condition. In Sec.~\ref{sec:4} we check WCCC for nonextremal BTZ black holes for the chiral gravity and Einstein gravity.  Finally in Sec.~\ref{sec:5} we summarize our results and conclude with some discussions on the issue of proving the third law and its implication to the holographic dual CFTs.

%%%%%%%%%%%%%%%%%     2 BTZ and var     %%%%%%%%%%%%%%%%%% %% 

\section{BTZ black hole and variational identities}
\label{sec:2}

BTZ black holes are topologically non-trival solutions to the 3-dimensional Einstein gravity as well as the topological massive gravity (TMG) \cite{Deser:1981wh,Deser:1982vy,Deser:2002iw}. In  fact, they are solutions to a quite general category of gravity theories with the name Mielke-Baekler (MB) model \cite{Mielke:1991nn,Baekler:1992ab} which also incorporates torsion, with Einstein gravity and TMG arise as special limits in its parameter space. In this section, we derive the variational identities and canonical energy for this model following Wald's formulation. 

In three dimensional spacetime, it is convenient to express the gravity theory in the first order formalism. The Lagrangian of a general chiral gravity with torsion, namely the MB model,  is as following: 
\be \label{actionMB}
 L \;=\; L_{\mbox{\tiny EC}}  \;+\;  L_{\L} \;+\; L_{\mbox{\tiny CS}}  \;+\; L_{\mbox{\tiny T}}  \;+\; L_{\mbox{\tiny M}} \;,
\ee
where
\bea 
L_{\mbox{\tiny EC}} &=& {1 \over \p} \, e^a \,\w\, R_a \;,  \label{actionEC} \\ 
L_{\L}  &=& - \, { \L \over 6 \p} \, \e_{abc} \, e^a \,\w\, e^b \,\w\, e^c \;,  \label{actionL}\\ 
L_{\mbox{\tiny CS}}  &=& -\, \tl \left( \om^a \,\w\, \dr \om_a  
\,+\, {1 \over 3} \,\e_{abc}\, \om^a \,\w\, \om^b \,\w\, \om^c \right)\;, \label{actionCS} \\ 
L_{\mbox{\tiny T}}  &=& { \tt \over 2 \p^2} \, e^a \,\w\, T_a \;, \label{actionT}
\eea
in which $\Lambda < 0$ is the cosmological constant, $\tl$ and $\tt$ are coupling constants, $L_{\mbox{\tiny EC}}$ is the Einstein-Cartan term, $L_{\L} $ is the cosmological constant term, $L_{\mbox{\tiny CS}} $ is the Chern-Simons (CS) terms for curvature, $L_{\mbox{\tiny T}} $ is a translational Chern-Simons term, and $L_{\mbox{\tiny M}}$ is the Lagrangian for the matter.  $T_a$ is the torsion 2-form defined by $T^a = \dr e^a + \om^a_{~b} \,\w\, e^b$ with $e^a$ the dreibeins. We have also defined the dual spin connection $\om^{a} $ and the dual curvature 2-form $R^{a} $ for simplicity:
\be 
\om^{a} \,=\, {1 \over 2} \,\e^a_{~bc} \,\om^{bc}\;,  \quad\quad\quad\quad 
R^{a} \,=\, {1 \over 2} \,\e^a_{~bc}\, R^{bc}\;, 
 \ee
Variations of the Lagrangian (\ref{actionMB}-\ref{actionT}) with respect to the dreibeins $e^a$ and the dual spin connections $\om^a$ give rise to the equations of motion $E_a^{(e)} = 0$ and $E_a^{(\om)} = 0$ with  
\bea
E_a^{(e)} &=& {1 \over \p} \left( R_a \,+\, {\tt \over \p} \, T_a 
\,-\, {\L \over 2} \, \e_{abc} \,e^b \,\w\, e^c \right) \;, \label{eom.e}\\
E_a^{(\om)}  &=&  {1 \over \p} \left( T_a \,-\, 2 \p \tl \, R_a 
\,+\, {\tt \over 2\p} \, \e_{abc} \, e^b \,\w\, e^c \right) \, \label{eom.om}
\eea
 for vanishing matter. For the case $ 1 + 2 \tt \tl  \neq 0\,$, the equations of motion are solved by 
\bea
T^a &=& { \cT \over \p } \,   \e^a_{~bc} \, e^b \,\w\, e^c \;, \label{T2} \\ 
R^a &=& - \, { \cR \over 2 \p^2 }   \,   \e^a_{~bc} \, e^b \,\w\, e^c \;, \label{R2}
\eea 
in which 
\be
\cT \,\equiv\, \frac{ -\tt \,+\, 2 \p^2 \L\tl  }{ 2 + 4 \tt \tl } \;,   \quad\quad 
\cR  \,\equiv\, - \, \frac{\tt^2 + \p^2 \L}{1 + 2 \tt \tl} \,. \label{calTR}
\ee

The MB model was originally proposed as a torsional generalization of TMG. It has a Poincar\'e gauge theory description, and there are propagating massive gravitons just like in TMG. We will be especially interested in three limits: 

(i) Einstein gravity (with negative cosmological constant). This could be approached by taking the limit $\,\tl \to 0\,$ and $\,\tt \to 0$\,.

(ii) Chiral gravity. The torsionless branch of the MB model, which is equivalent to TMG, could be obtained  by setting $\,{\cal T} = 0 \,$ according to (\ref{T2})\,. It was pointed out in \cite{Li:2008dq} that TMG is only well defined at the critical point in which the dual CFT becomes chiral. In our convention, the critical point is located at $\, \tl = - \, {1 /  ( 2 \p \sqrt{-\L} ) } \,$. Hence the chiral gravity is approached by setting $\,{\cal T} = 0 \,$ first and then taking the limit $\, \tl \to - \, {1 /  ( 2 \p \sqrt{-\L} ) } \,$.

(iii) Torsional chiral gravity. For the branch with non-vanishing torsion, we note from the Lagrangian (\ref{actionMB}-\ref{actionT}) that the torsion field $T_a$ could not be kinematic since there is no second order derivative of $\om^a$. The torsion field should just contribute to the interaction term in the linearized theory, while the propagators of the gravitons should not be changed compared with TMG. We then expect that the MB model also behaves well with no ghost at the critical point $\, \tl \to - \, {1 /  ( 2 \p \sqrt{-\L} ) } \,$. Note that by taking this limit first, we obtain $\,{\cal T} \to \p \sqrt{-\L} \,/ \,2\,$ hence the torsion field could not be vanishing. This is a different limit from the case (ii), and we refer it as the torsional chiral gravity.

 An interesting class of solutions to the equations (\ref{T2})(\ref{R2}) are the BTZ-like solutions with non-vanishing torsion \cite{Garcia:2003nm}. They are described by the following dreibeins 
 \be
e^0 \,=\, N \dr t\,,  \quad~~ 
e^1 \,=\, {\dr r \over N} \,, \quad~~
e^2 = r \left( \dr \f + N^{\f} \dr t \right) \, \label{eBTZ.1} \,,
\ee
and the dual spin connections 
\be \label{omBTZ.1}
\om^a \;=\; \omt^a \,+\, {\cT \over \p} \, e^a \,,
\ee
with the torsion free parts 
\be \label{omBTZ.2}
\omt^0 \,=\, N \dr \f\,, \quad~~ 
\omt^1 \,=\, - \, {N^\f \over N} \dr r \,,  \quad~~ 
\omt^2 \,=\, - \Lef \, r \dr t \,+\, r N^\f \dr \f  \,, 
\ee
in which
\be \label{eBTZ.2}
N^2 (r) \,=\, -M - \Lef \, r^2 + {J^2 \over 4 r^2 } \,,  \quad\quad
N^\f (r) \,=\,  - \, { J \over 2 r^2 } \,,
\ee
where $M$ and $J$ are constants corresponding to mass and angular momentum of BTZ black hole, respectively, for the case of Einstein's gravity, and 
\be \label{eBTZ.3}
\Lef \,\equiv\, - \,\frac{\cT^2 + \cR }{ \p^2 } \;.
\ee

Taking the torsion free limit $\cT \to 0$, the above solutions recover the usual BTZ black holes with $\Lef = \L$\,. 
The horizons are located at  
\be \label{rplus}
r_\pm^2 \;=\; {1 \over \, 2 \Lef} \left( - \, M \mp \sqrt{M^2 + \Lef J^2} \right) \,
\ee
(note that $\Lef < 0$ for asymptotic AdS solutions), and the angular velocity of the outer horizon is
\be
\OmH \;=\; {J \over 2 r_+^2} ={r_-\over r_+} \sqrt{-\Lef} \;.
\ee   
The black hole temperature is fixed by the vanishing of the conical singularity of the corresponding  Euclidean metric:  
\be
\TH \,=\, -\, \frac{\Lef \left( r_+^2 - r_-^2 \right) }{2 \p r_+} \,, 
\ee
and the surface gravity is $\kH = 2 \p \TH$\,. 

%%%%%%%%%%%%%%%%%           2.1  linear          %%%%%%%%%%%%%%%%%% %% 
\subsection{First order variations}

Wald's gedanken experiment to destroy a black hole begins with considering a general off-shell variation of the fields, which in principle incorporates all kinds of possible perturbations of a black hole, including throwing matter into it. From the variational identities one obtains general constrains obeyed by these perturbations.  

The first order variation of the Lagrangian (\ref{actionMB}-\ref{actionT}) gives rise to the equations of motion as well as a surface term: 
\be
\d L \;=\; \d e^a \,\w\, E_a^{(e)} \;+\; \d \om^a \,\w\, E_a^{(\om)} \;+\; \dr \Th (\f, \d \f) \,,
\ee
in which $\f = (e^a, \;\om^a)$, $E_a^{(e)}$ and $E_a^{(\om)}$ are given by (\ref{eom.e}) and (\ref{eom.om}). The surface term $\Th (\f, \d \f)$, called the symplectic potential, is evaluated to be 
\bea
\Th (\f, \d \f) &=&{ 1 \over \pi} \,\d \om^a \,\w\, e_a 
\;+\; {\tt \over 2 \p^2} \, \d e^a \,\w\, e_a \;-\; \tl \, \d \om^a \,\w\, \om_a \,.  
\eea
In Wald's approach, the space of field configurations is the phase space of the theory, and the variation $\d \f \equiv (d \f / d \l) |_{\l = 0}$ is the phase space flow vector associated with a 1-parameter family of field configurations $\f(\l)$. For a 2-parameter family of field configurations $\f(\l_1, \l_2)$, one could define the symplectic current 
\be
\Om (\f, \d_1 \f, \d_2 \f) \,=\, \d_1 \Th(\f, \d_2 \f) \,-\, \d_2 \Th(\f, \d_1 \f) \,,
\ee
in which $\d_1, \,\d_2$ denote derivatives with respect to parameters $\l_1, \, \l_2$: 
\be
\d_1 = {\pl \over \pl\l_1}\bigg |_{\l = 0}\,, \quad\quad\quad  \d_2 = {\pl \over \pl\l_2}\bigg |_{\l = 0}\,. 
\ee  
One can show that the symplectic current is conserved when the linearized equations of motion are satisfied:
\be
\dr \Om \,=\, 0\,.  
\ee

The Noether current 2-form associated with a vector field $\xi$ is defined by
\be  \label{noethercurrent1}
j_\xi \;=\; \Th(\f, {\cal L}_\xi \f) - \in L\,,
\ee
in which $\,\in L$ represents the interior derivative which contracts $\,\xi^\m$ into the first index of the 3-form $L$.
Then, $j_\xi$ could be written in the form
\be \label{noethercurrent2}
j_{\xi} \;=\; \dr Q_\xi \,+\, C_\xi \,,
\ee
in which the Noether charge $Q_\xi$ and the constraints $C_\xi$ are given by 
\bea
Q_\xi &=& {1 \over \p} \, ( \in \om^a ) \,\w\, e_a \;+\; {\tt \over 2\p^2} \, (\in e^a) \,\w\, e_a 
\,-\, \tl \,(\in \om^a ) \,\w\, \om_a \,,  \\ ~\nn \\ 
C_\xi &=& -(\in e^a)  \,\w\, E_a^{(e)} \,-\, (\in \om^a )  \,\w\, E_a^{(\om)} \,. \label{Cxi}
\eea

Suppose the field configuration is a family of asymptotic AdS spacetime. Variation of equations (\ref{noethercurrent1})(\ref{noethercurrent2}) gives rise to the following linear variational identity after integrating over an achronal hypersurface $\S$: 
\be \label{1var}
\int_{\pl \S} \d Q_\xi \,-\, \in \Th(\f, \d \f)  \;\,=\;\, \int_{\S} \Om(\f, \d \f, {\cal L}_\xi \f) 
- \int_{\S} \d C_\xi - \int_{\S} \in ( E  \d \f )\,.
\ee
The first term on the right hand side is recognized as the variation of the Hamiltonian $ h_{\xi}$ associated with the diffeomorphism generated by the vector field $\xi$\,
\be
\d h_\xi \;=\; \int_{\S} \Om(\f, \d \f, {\cal L}_\xi \f)  \,.
\ee
Note that $\d h_\xi$ (or the first term on the R.H.S. of \eq{1var}) vanishes if $\xi$ is a Killing field, i.e., ${\cal L}_\xi \f=0$. If the $\f$ is on-shell so that $E_a = 0$, then the last two terms on the R.H.S. of \eq{1var} also vanish. This then motivates the following definition of the conserved ADM quantity $H_\xi$ conjugate to the Killing field $\xi$ for an on-shell $\f$ \cite{Wald:1993nt}\cite{Iyer:1994ys}:
\be \label{WaldCharge}
\d H_\xi \;=\; \int_{\infty} \d Q_\xi \,-\, \in \Th(\f, \d \f) \,,
\ee
where $\int_{\infty}$ is the integration over the circle at spatial infinity.  For a black hole solution, the boundary of $\Sigma$ contains also horizon as the ``inner boundary" besides the ``outer boundary" at spatial infinity, then there will be contribution to the L.H.S. of  \eq{1var} from the ``inner boundary" as well  (i.e., the area law term for the Einstein gravity). Combining all the above, \eq{1var} finally yields the first law of black hole mechanics/thermodynamics.

For the timelike Killing field $\partial/\partial t$ and the rotational Killing field $\partial/\partial \vphi$\,, the above integration gives rise to the variation of the total mass $\cal M$ and the total angular momentum $\cal J$\,, respectively. For the BTZ-like black holes (\ref{eBTZ.1}-\ref{omBTZ.2}), it could be evaluated that \cite{Ning:2018gfm} 
\bea
{\cal{M}}  &=&  M \,-\, 2 \tl \left( {\cal{T}}  M \,+\, \p \Lef J \right) \;,  \label{energy} \\
 {\cal{J}} &=&   J ~ \,+\, 2 \tl \left( \p  M \,-\, {\cal{T}}  J \right) \,.  \label{angularmomentum}
\eea

For the case that the equations of motion are satisfied and $\xi$ is a Killing field, the linear variational identity (\ref{1var}) yields
\be \label{1varII}
\int_{\pl \S} \d Q_\xi \,-\, \in \Th(\f, \d \f)  \;\,=\;\, - \int_{\S} \d C_\xi \,.
\ee
For nonextremal black holes, the boundaries include the infinity as well as the bifurcation surface $B$\,. If $\xi$ is the horizon Killing field $\xi^a = \partial/\partial t + \OmH \partial/ \partial \vphi$\,, the boundary integral over infinity is given by 
\be \label{infinity1}
\int_{\infty} \d Q_\xi \,-\, \in \Th(\f, \d \f)  \;\,=\;\, \d {\cal M} - \OmH \d {\cal J}\,. 
\ee
For Einstein's theory of gravity, the boundary contribution from the bifurcation surface $B$ turns out to be proportional to the variation of the Bekenstein-Hawking entropy \cite{Wald:1993nt}\cite{Iyer:1994ys}: 
\be \label{BHentropy}
\int_{B} \d Q_\xi \,-\, \in \Th(\f, \d \f)   \;\,=\;\,   \TH \,\d S \,,
\ee
in which $S = A_B / 4\,$ where $A_B$ is the area of the bifurcation surface. We will take the above equation as a rightful definition of the modified black hole entropy in the MB model so that the first law of the black hole thermaldynamics still holds but with $S$ Wald's generalized black hole entropy. It has been evaluated for the  BTZ-like black holes that \cite{Ning:2018gfm} 
\be \label{entropy}
S \;=\; 4\p r_+ \,-\, 8 \p \tl \left( {\cal T} r_+  -  \pi \sqrt{-\Lef}\,  r_- \right)\;.
\ee
The equation (\ref{1varII}) then takes the form
\be \label{1varIII}
\d {\cal M} - \OmH \d {\cal J}  - \TH \,\d S  \;=\; - \int_{\S} \d C_\xi \,.
\ee
We will consider the special situation that the perturbation vanishes near the internal boundary of the surface $\S$, then equation (\ref{1varIII}) with $\d S = 0$ would hold for both extremal and non-extremal black holes. Noting (\ref{energy})(\ref{angularmomentum}) and $\d S = 0$\,, (\ref{1varIII}) turns out to be
 \be \label{1varIV}
\left( 1 \,-\, 2 \tl {\cal T} \,-\, 2 \pi \tl \OmH \right) \left( \d M \,-\, \OmH \d J \right) 
\,-\, 2 \pi \tl \Lef \left( { r_+^2 - r_-^2  \over r_+^2 }  \right) \d J
\;=\; - \int_{\S} \d C_\xi \,.
\ee 
for BTZ-like black holes in the MB model. 

The equations (\ref{1varIII})(\ref{1varIV}) are derived from the Lagrangian without matter. It might be puzzling that the vacuum configuration could be perturbed without matter; however, this is physically possible since there are gravitational waves in the MB model with general couplings. In general, $\d M$ and $\d J$ should be understood as variations allowed mathematically in the parameter space, rather than consequences of certain physical evolutions. On the other hand, since we didn't enforce the linearized equations of motion to be satisfied, it should be expected that these equations could also be used for considering perturbed configurations due to matter contribution\footnote{For the matter field, we impose Dirichlet condition on asymptotic AdS boundary, as conventionally used in AdS/CFT dictionary for black holes dual to thermal states in CFT. This choice will not affect the argument for WCCC as we only care about the matter that falls in.}. The right hand side of (\ref{1varIV}) would be related to the energy-momentum tensor of the matter. To see this explicitly, we first define the ``energy-momentum 2-form'' $\S_a$\, and ``spin current 2-form'' $\t_a$\, as follows:
\be 
\S_a \;\equiv\;  {\d L_{\mbox{\tiny M}} \over \d e^a } \;, \quad\quad\quad~~ 
\t_a \;\equiv\;  {\d L_{\mbox{\tiny M}} \over \d \om^a }  \label{source} \,.
\ee
The equations of motion with matter would be
\be \label{onshellEOM}
E_a^{(e)} \;=\; - \, \S_a\;, \quad\quad\quad~~
E_a^{(\om)} \;=\;  - \, \t_a \,.
\ee
Since $\S_a = \t_a = 0$ in the background spacetime, from (\ref{Cxi}) we get
\be \label{dC}
\d C_\xi \;\,=\;\, (\in e^a)  \,\w\, \d \S_a \,+\, (\in \om^a )  \,\w\, \d \t_a \;.
\ee 
$\S_a$ should be related to the conserved canonical energy-momentum tensor $\S_a^{~\m} $ defined by
\be \label{canonicalSigma}
\sqrt{-g} \;\S_a^{~\m}  \;\equiv\; {\pl {\cal L}  \over \,\pl e^a_{~\m} \,}  
\;=\; e_a^{~\m} {\cal L} \,-\, {\pl {\cal L}  \over \,\pl (\pl_\m \psi) \, } \, D_a \psi\,,
\ee
in which $\psi$ is the matter field, $\cal L$ is the Lagrangian density of the matter related to $L_{\mbox{\tiny M}}$ by $L_{\mbox{\tiny M}} = {\cal L}\, d^3 x$, and $D_a$ is the covariant derivative defined by $D_a = e_a^{~\m} (\pl_\m + \om_\m^{~\,bc} f_{cb} )$ where $f_{ab}$ are the representations of the generators of the Lorentz group associated with $\psi$. From (\ref{canonicalSigma}) we obtain 
\be \label{stress1form}
\S_a \;=\;  {1 \over 2} \, \e_{\m\n\l} \,\S_a^{~\l} \,dx^\m \w\, dx^\n \,.
\ee 
Note that 
\be
 \e_{\m\n\l} \;=\; - \,3 k_{[\m} {\hat \e}_{\n\l]} \,,
\ee
in which $k^\m$ is the future-directed normal vector to the horizon, and $\hat \e$ is the volume element on the horizon. The first term on the right hand side of (\ref{dC}) then turns out to be
\be
 (\in e^a)  \,\w\, \d \S_a   \;=\;  - \,  \xi_\m k_{\n}   \d \S^{\m\n} \sqrt{-\g} \,d^2 x\;,
\ee
as $\,\xi^\m \propto k^\m$\,, the contribution of this term to the right hand side of equation (\ref{1varIV}) is non-negative if and only if the null energy condition of matter energy-momentum tensor $\,\d \S_{\m\n}$\, is satisfied:
\be
k^\m k^\n \d \S_{\m\n} \,\ge\, 0 \,. 
\ee

For the second term on the right hand side of (\ref{dC}), our ``spin current 2-form'' $\t_a$\, is related to the  canonical spin angular momentum tensor $\t_{ab}^{~~\,\m}\,$ defined by 
\be \label{canonicalSpin}
\sqrt{-g} \; \t_{ab}^{~~\,\m} \;\equiv\; {\pl {\cal L}  \over \pl \om_\m^{~\,ab}} 
\;=\; - \, {\pl {\cal L}  \over  \pl (\pl_\m \psi )} \, f_{ab} \,\psi \,,
\ee
Comparing (\ref{canonicalSpin}) with (\ref{source}), we obtain  
\be \label{spin1form}
\t_a \;=\; - \, {1 \over 2} \, \e_a^{~bc} \, \e_{\m\n\l} \, \t_{bc}^{~\;\l} \,dx^\m \,\w\, dx^\n \,,
\ee
hence the second term on the right hand side of (\ref{dC}) is reduced to 
\be \label{spinEC1}
(\in \om^a)  \,\w\, \d \t_a  \;=\; - \, (\xi^\s \om^{ab}_{~~\;\s})\, k_{\l}\, \d \t_{ab}^{~~\l} \sqrt{-\g} \, d^2 x\;. 
\ee
For axially symmetric stationary black holes, in general we have  \cite{Ning:2018gfm}  
\be \label{zeta.omega}
\in \om^a \,|_{\cal H} \;=\; - \, {1 \over 2}\, \kH \, \e^a_{~bc} \, n^{bc}
 \,+\, \in K^a  \,|_{\cal H} \;,
\ee
in which $n^{ab}$ is the binormal to the horizon and $K^a$ is the dual contorsion 1-form defined by $T^a \,=\, \e^a_{~bc} \, K^b \,\w\, e^c \,$, satisfying the identity $ \om^a \,=\, \omt^a \,+\, K^a$. For BTZ-like black holes, (\ref{omBTZ.1}) gives   
\be \label{contorsion}
K^a \;=\; {{\,\cal T} \over \,\p} \,e^a \,.
\ee
Using (\ref{zeta.omega})(\ref{contorsion}), equation (\ref{spinEC1}) turns out to be
\be \label{dC2}
(\in \om^a)  \,\w\, \d \t_a  \;=\; 
\left(  \kH n_{\m\n} \,+\,  
 \frac{\,\cal T}{\,\p} \, \e_{\m\n}^{~~\;\s} \xi_\s \right) k_\l  \d \t^{\m\n\l}  \sqrt{-\g}  \, d^2 x \,. 
\ee
The first term on the right hand side is vanishing for extremal black holes.  We note that the sign of the second term could not be determined for torsional chiral gravity unless the spin angular momentum tensor satisfies $\, \e_{\m\n}^{~~\;\s} k_\s k_\l  \d \t^{\m\n\l} \,\ge\, 0 \, $, of which the physical meaning is not clear yet for us. 

Combining all the results above, we obtain the following linear variational identity for the BTZ-like black holes in the MB model, with the additional assumption that the perturbation $\d \f$ vanishes near the internal boundary of $\S$
 \bea \label{1varV}
&&  \d {\cal M} \,-\, \OmH \d {\cal J}  \nn  \\
 &=& \left( 1 \,-\, 2 \tl {\cal T} \,-\, 2 \pi \tl \OmH \right) \left( \d M \,-\, \OmH  \d J \right) 
\,-\, 2 \pi \tl \Lef \left( { r_+^2 - r_-^2  \over r_+^2 }  \right) \d J    \nn \\
&=&  \int_{\S} d^2 x  \sqrt{-\g} \left\{ \xi_\m k_{\n}   \d \S^{\m\n}  
-  \left( \kH n_{\m\n} + \frac{\,\cal T}{\,\p} \,  \e_{\m\n}^{~~\;\s} \xi_\s \right) k_\l  \d \t^{\m\n\l} \right\} \,. 
\eea
For extremal BTZ black holes with $\,\kH = 0\,$ and $\,r_+ = r_-\,$\,, the above identity takes the following simpler form: 
 \bea \label{1varVI}
  \d {\cal M} \,-\, \OmH \d {\cal J}  
 &=& \left( 1 \,-\, 2 \tl {\cal T} \,-\, 2 \pi \tl \sqrt{- \Lef} \right) \left( \d M \,-\, \sqrt{- \Lef} \, \d J \right)   \nn \\
&=&  \int_{\S} d^2 x  \sqrt{-\g} \left\{ \xi_\m k_{\n}   \d \S^{\m\n}  
- \frac{\,\cal T}{\,\p} \,  \e_{\m\n}^{~~\;\s} \xi_\s  k_\l  \d \t^{\m\n\l} \right\} \,. 
\eea

%%%%%%%%%%%%%%%%%           2.2  2nd order          %%%%%%%%%%%%%%%%%% %% 

\subsection{Second order variations}

As pointed out in \cite{Sorce:2017dst}, for near-extremal black holes it is in general not sufficient to consider just the linear order variation due to Hubeny-type violations. We therefore construct further the second order variational identity. A second variation of equation (\ref{1var}) gives rise to
\be \label{2nd1}
{\cal E}_\S(\f; \d \f) \;=\; \int_{\pl \S} \left[ \d^2 Q_\xi \,-\, \in \d \Th(\f, \d \f) \right] \,+\, 
\int_{\S} \d^2 C_\xi  \,+\,  \int_{\S} \in \left(\d E \,\w\, \d \f \right) \,, 
\ee
in which 
\be
{\cal E}_\S(\f; \d \f) \;\equiv\;  \int_{\S} \Om(\f, \d \f, {\cal L}_\xi \d\f) 
\ee
is Wald's canonical energy of the off-shell perturbation $\d \f$\, on $\S$\,. For the case that the background $\f$ is a stationary black hole solution and $\xi$ is the horizon Killing field, the boundary contribution from infinity is simply
\be \label{2nd2}
\int_{\infty} \d^2 Q_\xi \,-\, \in \d \Th(\f, \d \f)  \;\,=\;\, \d^2 {\cal M} \,-\, \OmH \d^2 {\cal J} 
\ee
according to (\ref{infinity1}). The contribution from interior boundary would be vanishing if there's no perturbation in its neighborhood, as supposed before.  Then equation (\ref{2nd1}) turns out to be 
\be  \label{2ndBig}
\d^2 {\cal M} \,-\, \OmH \d^2 {\cal J} \;\;=\;\; 
{\cal E}_{\S} (\f; \d \f) \,-\,  \int_{\S} \in \left(\d E \,\w\, \d \f \right) 
\,-\,  \int_{\S} \d^2 C_\xi  \;.
\ee

Noting (\ref{onshellEOM})(\ref{stress1form})(\ref{spin1form}), the integrand of the second term on the right hand side is evaluated to be
\bea \label{Term2}
\in \left(\d E \,\w\, \d \f \right)  &\equiv&  
\in \left(\d E_a^{(e)} \,\w\, \d e^a \;+\; \d E_a^{(\om)} \,\w\, \d \om^a \right)   \nn \\
&=& \xi^\t \, \Xi_{[\m\n\t]} \,dx^\m \,\w\ dx^\n \,,  \label{pullback0}
\eea
in which 
\be
\Xi_{\m\n\t}  \;=\; -\; {3 \over 2} \,\e_{\m\n\l} \left( \d \S_a^{~\,\l} \, \d e^a_{~\,\t} 
\,+\, \d \t_{ab}^{~~\l} \, \d \om^{ab}_{~~\;\t} \right) \,.
\ee
Since $\xi$ is tangent to the horizon, the pullback of (\ref{Term2}) to the horizon vanishes, hence this term gives no contribution. From (\ref{Cxi}), it turns out that 
\be  \label{d2C0}
\d^2 C_\xi \;=\; \d^2 \left\{ - \, d^2 x \sqrt{-\g} \left[ \xi_\m k_{\n}  \S^{\m\n}  
  \,-\,  \left( \kH n_{\m\n} + \frac{\,\cal T} {\,\p} \, \e_{\m\n}^{~~\;\s} \xi_\s \right) k_\l  \t^{\m\n\l} \right] \right\}  \,. 
\ee
Substituting the above expression into (\ref{2ndBig}) leads to the following identity for the second order variation: 
\bea \label{2ndVarId}
\d^2 {\cal M} \,-\, \OmH \d^2 {\cal J}  &=&  {\cal E}_{\S} (\f; \d \f)  \nn \\
&&  \,+\, \d^2 \int_{\S}   d^2 x \sqrt{-\g} 
\left\{  \xi_\m k_{\n} \S^{\m\n}  
  \,-\,  \left( \kH n_{\m\n} + \frac{\,\cal T} {\,\p} \, \e_{\m\n}^{~~\,\s} \xi_\s \right) k_\l \t^{\m\n\l}  \right\} \,.  
\nn \\
\eea

%%%%%%%%%%%%%%%%%    3  extre    %%%%%%%%%%%%%%%%%%
        
\section{Gedanken experiment to destroy an extremal BTZ}
\label{sec:3}

We now consider our gedanken experiment to destroy a BTZ black hole along the line of Wald's proposals \cite{Wald:I,Sorce:2017dst}.  In this section, we will deal with an extremal BTZ black hole with mass parameter $M$ and angular momentum parameter $J$.  We wish to see if a naked singularity can be made via throwing matter into the extremal black hole.  Without losing generality, we take our gravity theory as MB model, and then discuss its three limits, torsional chiral gravity, chiral gravity and three-dimensional Einstein gravity.

Considering a 1-parameter family of solutions $\f(\l)$, $\f_0=\f(0)$ is an extremal BTZ black hole, which is a vacuum solution in MB model.  The existence of event horizon is determined by a function,
\be \label{eq:f-func}
   f(\l) = M(\l)^2 + \Lef   J(\l)^2 \,,
\ee
If $f(\l)\geq 0$, the spacetime is a BTZ black hole.  If $f(\l)<0$, it is a naked conical singularity and WCCC is violated.  We now consider perturbations to the extremal black hole $\f_0$.  Then, to the first order in $\l$, we have
\be 
   f(\l) = 2\l \sqrt{-\,\Lef \,}  \vert J \vert \lrbrk{\d M -\sqrt{-\,\Lef \,}  \d J} + \cO(\l^2)\,,
\ee
where we have used the extremal condition $M = \sqrt{-\,\Lef \,}  \vert J \vert$ to eliminate $M$.  It is then evident that if $\d M < \sqrt{-\,\Lef \,}  \d J$, $f(\l)$ can be negative.

We would like to see whether this sort of violation of WCCC is possible if we throw matter into the BTZ black hole in a certain way.  Let $\S_0$ be an asymptotically AdS hypersurface which extends from the future horizon to the spatial infinity.  We consider a perturbation $\d \f$ whose initial data for both fields $\delta e^a$ and $\d \om^a$ on $\S_0$ vanish in the neighborhood of the intersection between $\S_0$ and the horizon. We assume that the initial data for the matter sources $\d \S_{\mu \nu}$ and $\d \tau^{\mu \nu \l}$ also vanish in this neighborhood, and, only exist in a compact region of $\S_0$.  That is, we consider perturbations whose effects at sufficiently early times are negligibly small.  To simplify the discussion, we only consider the case where, as we evolve the perturbation, all of the matter will fall through the horizon.    Therefore, the whole evolutions of the matter source $\d \S_{\mu \nu}$ and $\d \tau^{\mu \nu \l}$ stay in a shaded region as shown in Fig.~\ref{fig:extre-BTZ}.  As matter falls in, we further define a hypersurface $\S$ in the following way---it starts on the future horizon in the region where the perturbation vanishes and extends along the future horizon till all matter falls into the horizon; then it becomes spacelike, approaches the spatial infinity and becomes asymptotically AdS. We denote the horizon portion of $\S$ as $\cH$, and the spatial portion as $\S_1$.

\begin{figure}[hbt!] 
  \centering
  \captionsetup{width=5.5in}
  \includegraphics[width=5.5in]{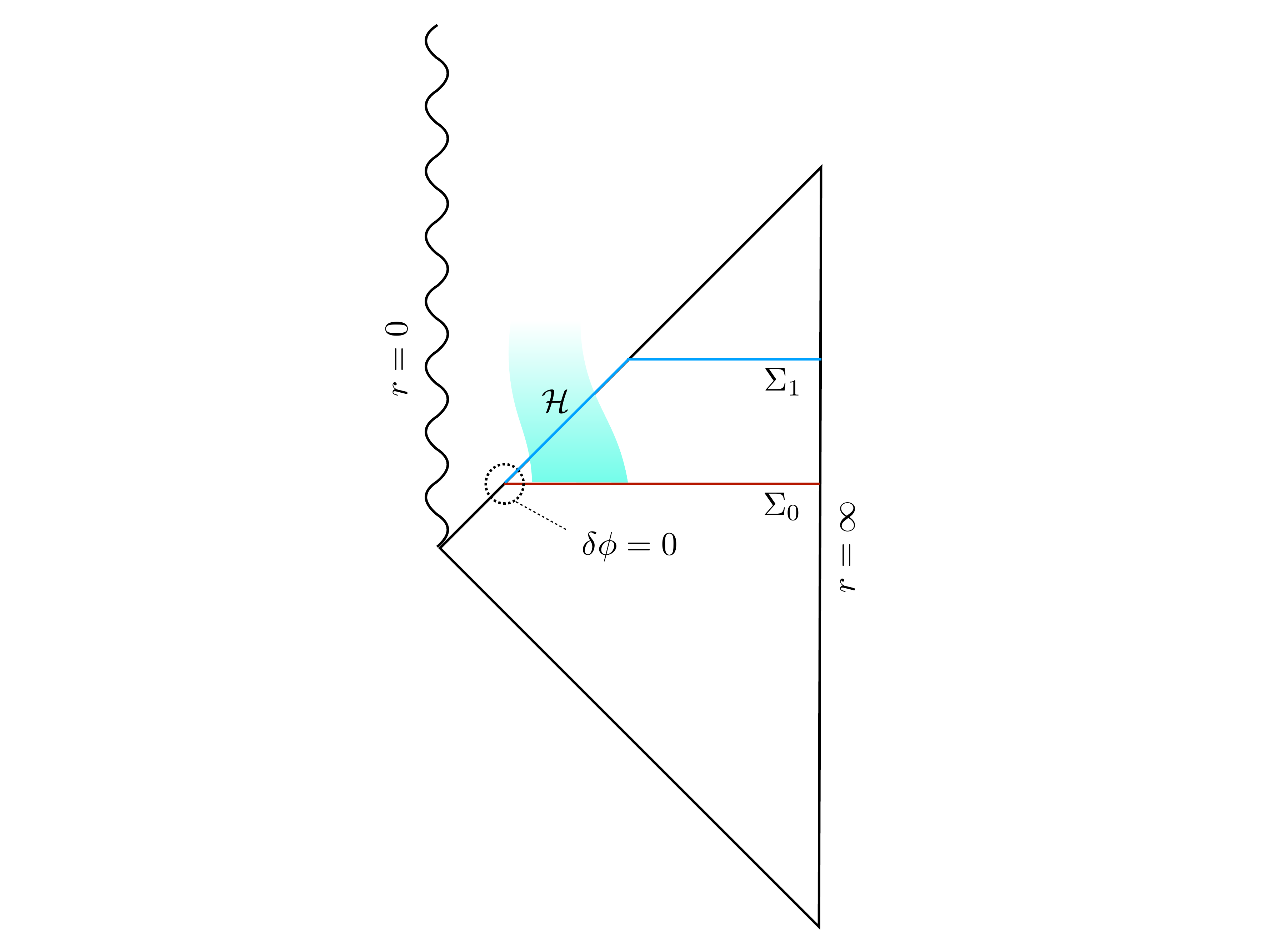}
  \caption{Carter-Penrose diagram of an extremal BTZ black hole.  The shaded region consists of the falling matter which all goes into the black hole.  The perturbation $\d \f$ vanishes in a neighborhood of $\S_0 \cap \cH$.}
  \label{fig:extre-BTZ}
\end{figure}

We now use the linear variational identity with vanishing inner boundary contributions~\eqref{1varVI} for this choice of $\S$.  As we will show later, this identity constrains the sign of $f(\l)$.  We notice that in e.q.~\eqref{1varVI}, the integral in the second line is not positive definite due to the spin angular momentum term and its coupling to torsion.   That is, in torsional chiral gravity, whether WCCC can hold depends on an additional relation between the spin angular momentum and the torsion.  The physical origin of this additional information needed is unclear, and is beyond our scope of this paper. We will leave it to a future work.   In the torsionless limit $\,{\cal T} \to 0$\,\,, however, this integral would be non-negative as long as the null energy condition is satisfied.  From now on, we will focus on this limit, and assume the falling matter satisfies the null energy condition.  Then $f(\l)$ is non-negative only if the constant factor on the rhs of the first line of e.q.~\eqref{1varVI} is non-negative,
\be \label{eq:ex-btz-WCCC}
1- 2 \th_{\rm L} \cT -2 \pi \th_{\rm L}\sqrt{-\,\Lef \,}  \geq 0\,.
\ee

For chiral gravity, we choose $\th_L = -1/(2\pi \sqrt{-\L})$, and send $\cT \to 0$.   The inequality~\eqref{eq:ex-btz-WCCC} is then satisfied.  Therefore extremal BTZ black hole in chiral gravity cannot be destroyed in our experiment, and WCCC is preserved.

For 3-dimensional Einstein gravity with a negative cosmological constant, both torsion and Chern-Simons interaction vanish, thus we set $\th_{\rm L} \to 0$ and $\th_{\rm T} \to 0$.  The inequality~\eqref{eq:ex-btz-WCCC} is then satisfied.  Consequently, extremal BTZ black hole in 3-dimensional Einstein gravity cannot be destroyed, leaving WCCC preserved.

%%%%%%%%%%%%%%%%%    4 near-extre   %%%%%%%%%%%%%%%%%%%%

\section{Gedanken experiment to destroy a near-extremal BTZ} 
\label{sec:4}

For extremal BTZ black holes, we have found that WCCC can be violated in the presence of torsion.  With torsion being turned off, we have seen that WCCC is preserved in both chiral gravity and 3-dimensional Einstein gravity, provided that the matter obeys the null energy condition.  In 4-dimensional Einstein gravity, Hubeny~\cite{Hubeny:1998ga} proposed that violations of WCCC might be possible if one threw matter into a near-extremal black hole in an appropriate manner.  In order to examine whether Hubeny-type violations can truly happen, one has to calculate the energy and momentum of the matter beyond the linear order.  In this section, we will examine the Hubeny-type violations for a near-extremal BTZ black hole in chiral gravity and 3-dimensional Einstein gravity respectively.

As shown in Fig.~\ref{fig:nonextre-BTZ}, we  make similar choices of $\S_0$ and $\S$ like those for the extremal BTZ case.  The only difference is that, the two hypersurfaces now terminate at the bifurcation surface $B$.  We further assume that the second order perturbation $\d^2\f$ for both fields $\d e^a$ and $\d \om^a$ also vanishes in a neighborhood of $B$.  Again, we simplify our discussions by restricting to the case where all matter falls into the black hole.  We will also make the following additional assumption: 
\begin{quote}
{\it Assumption}: The non-extremal BTZ black hole is linearly stable to perturbations, i.e., any source-free linear perturbation $\d \f$ approaches a perturbation $\d \f^{\rm BTZ}$ towards another BTZ  black hole at sufficiently late times.
\end{quote}

Although our perturbations are not source-free in general, we will only apply the above assumption on the late-time spatial surface $\S_1$ long after all of the matter has fallen in the the black hole. We emphasize that this linear stability assumption does not indicate WCCC which we wish to prove. This is because a finite perturbation is needed to overspin a non-extremal black hole, while a linear perturbation can always be scaled down to a infinite small one. Hence the linear instability of a non-extremal BTZ black hole should be independent of the instability associated with overspinning, i.e., a linearly stable non-extremal BTZ black hole could possibly be overspun by a finite perturbation, just like the situation for the Kerr-Newman black hole in \cite{Sorce:2017dst}.

\begin{figure}[hbt!] 
  \centering
  \captionsetup{width=5.5in}
  \includegraphics[width=5.5in]{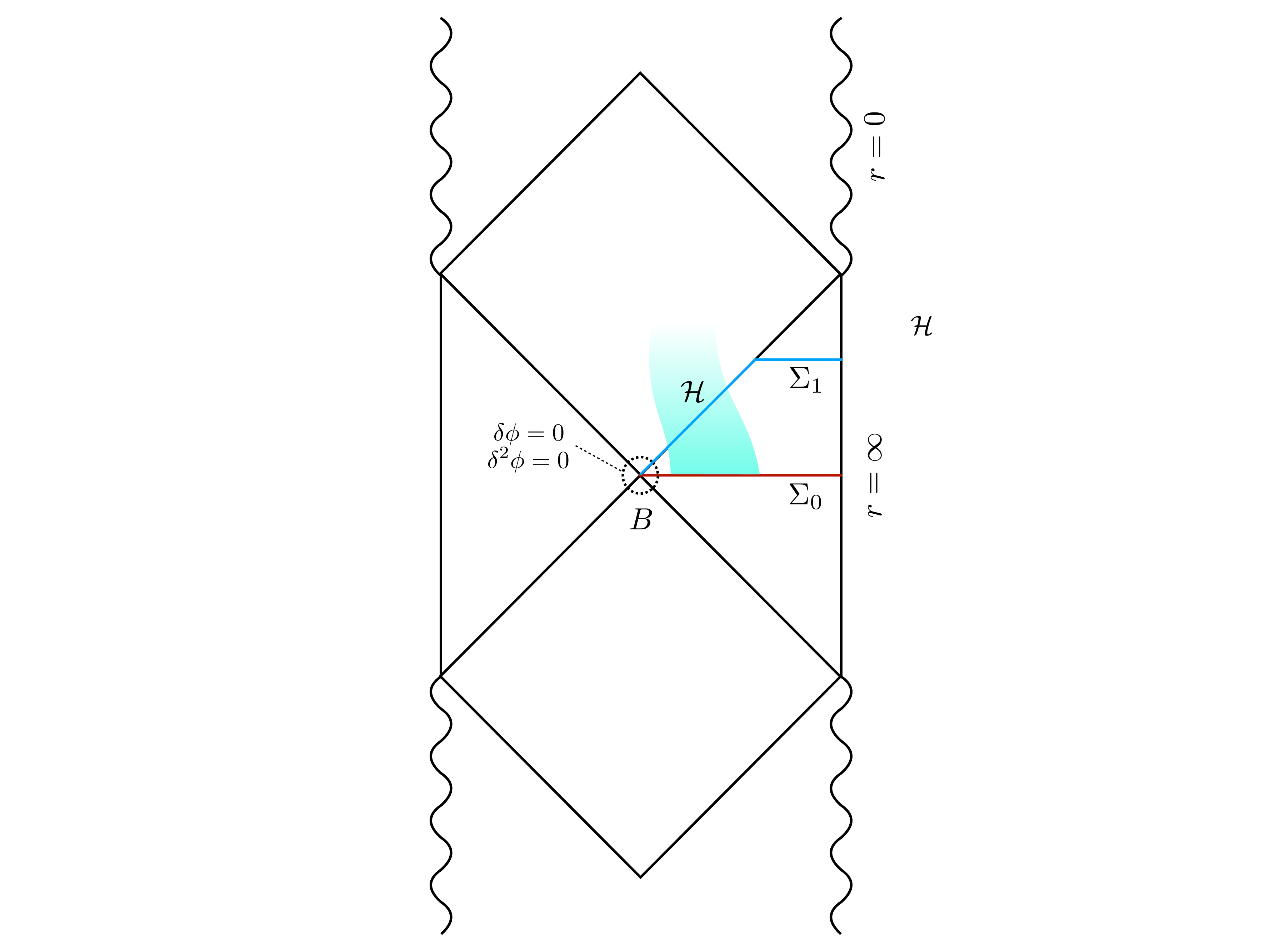}
  \caption{Carter-Penrose diagram of a near-extremal BTZ black hole.  The shaded region consists of the falling matter which all goes into the black hole.  The perturbation $\d \f$ and $\d^2 \f$ vanishes in a neighborhood of $B$.}
  \label{fig:nonextre-BTZ}
\end{figure}

\subsection{Chiral gravity}

We now consider our thought experiment to destroy a near-extremal BTZ black hole $(M,J)$ in chiral gravity for which ${\cal T}=0$ and $\tl = - \, {1 \over   2 \p \sqrt{-\L}  }$. Thus, using (\ref{energy}-\ref{angularmomentum}) it is straightforward to see 
\be
\d \cM - \Om_{\rm H}\d \cJ = \lrbrk{1+\frac{\Omega_H}{\sqrt{-\Lambda}}} \lrbrk{\d M - \sqrt{-\L} \d J}\,,
\ee
and the first law of black hole thermodynamics yields
\be
T_{\rm H}\d S =  \d \cM - \Om_{\rm H}\d \cJ\,.
\ee
where the black hole entropy is 
given by~\cite{Ning:2018gfm}
\be
S  = 4\pi \lrbrk{r_+ - r_-}\,.
\ee

Recall (\ref{1varV}), the null energy condition for the falling matter yields the first order relation that
\be
\d M \geq \sqrt{-\L} \d J \,.
\ee
Assuming the first order perturbation has been optimally done, i.e. $\d S = 0$, such that
\be
\d M = \sqrt{-\L} \d J \,.
\ee
For some constant entropy $S$, we can then plot the line of constant entropy in the parameter space of BTZ black holes, which is shown in Fig.~\ref{fig:hubeny_BTZ}.

\begin{figure}[hbt!] 
  \centering
  \captionsetup{width=5.5in}
  \includegraphics[width=4.5in]{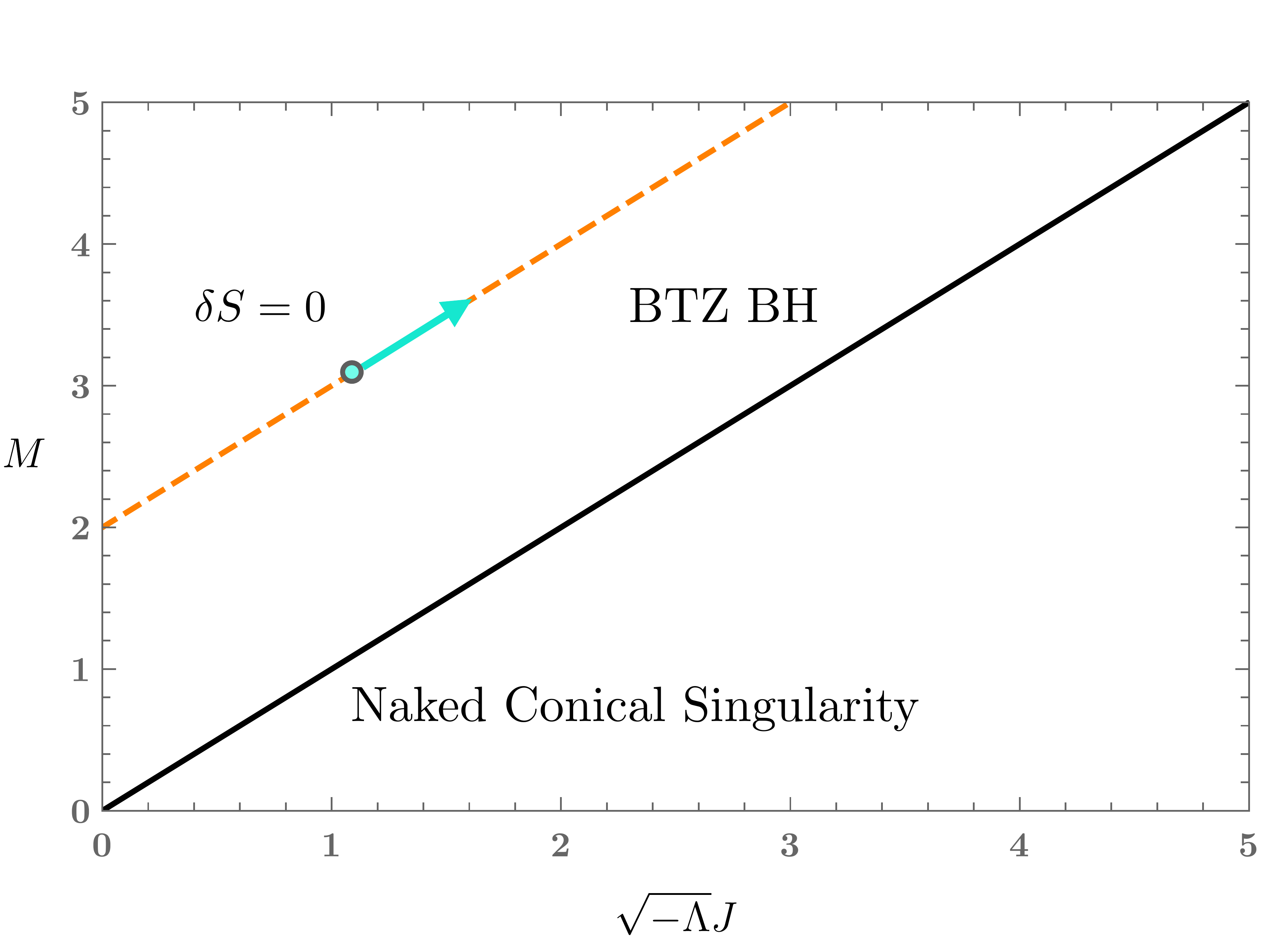}
  \caption{The parameter space of BTZ black holes in chiral gravity.  The black solid line corresponds to extremal BTZ black holes.  Any point above this line corresponds to a non-extremal BTZ black hole, while any point below the line is a naked conical singularity.  The orange dashed line is one of the lines of constant entropy, which is parallel to the line for extremal BTZ black holes.  Starting with some point on the constant entropy line, any tangent vector will always be parallel to the extremal BTZ line.  That is, there is no Hubeny-type violation that can overspin a near-extremal BTZ black hole in chiral gravity.}
  \label{fig:hubeny_BTZ}
\end{figure}

We are now ready to discuss our experiment to destroy the near-extremal BTZ black hole.  
Starting from a point $(M_0,J_0)$ in the parameter space, after a perturbation of the spacetime as induced by falling matter, we will always arrive at another point $(M_1,J_1)$.  At the linear order, the change from one point to another will correspond to a tangent vector in the parameter space.  For any $S$,
the line of constant entropy is given by
\be
M = \lrbrk{\sqrt{-\L}} J -\frac{\L}{16 \pi^2}S^2 \,.
\ee
The slope of the constant entropy line is then equal to that of the line representing extremal BTZ black holes.
Since the tangent to the constant entropy line is a lower bound to all physically-realizable perturbations, a non-extremal BTZ black hole will at most be perturbed to another BTZ black hole with the same entropy.  
There is no Hubeny-type violation of weak cosmic censorship for the BTZ black hole in three-dimensional chiral gravity, thus WCCC is preserved.

\subsection{Einstein gravity}

The discussions above can be applied to the BTZ black holes in 3-dimensional Einstein gravity as well,  for which we turn off both torsion and Chern-Simons interactions in MB model. In this theory, the linear variational identity is given by 
\be
\d \cM - \Om_{\rm H} \d \cJ = \d M- \Om_{\rm H} \d J\,.
\ee
Given the material null energy condition, we similarly find that 
\be
\d M- \Om_{\rm H} \d J \geq 0\,.
\ee
Once a first order perturbation is optimally done by choosing $\d M= \Om_{\rm H} \d J$, according to the first law of black hole thermodynamics, we will also find a lower bound for all perturbations given by $\d S=0$.  In Einstein gravity, the entropy of the BTZ black hole is $S = 4\pi r_+$, and the curve of constant entropy is given by
\be
M = \frac{4\pi^2}{S^2}J^2 - \frac{\L}{16\pi^2}S^2 \,.
\ee
We plot one of such curves in Fig.~\ref{fig:hubeny-BTZ-Einstein}.

\begin{figure}[hbt!]
  \centering
  \captionsetup{width=5.5in}
  \includegraphics[width=4.5in]{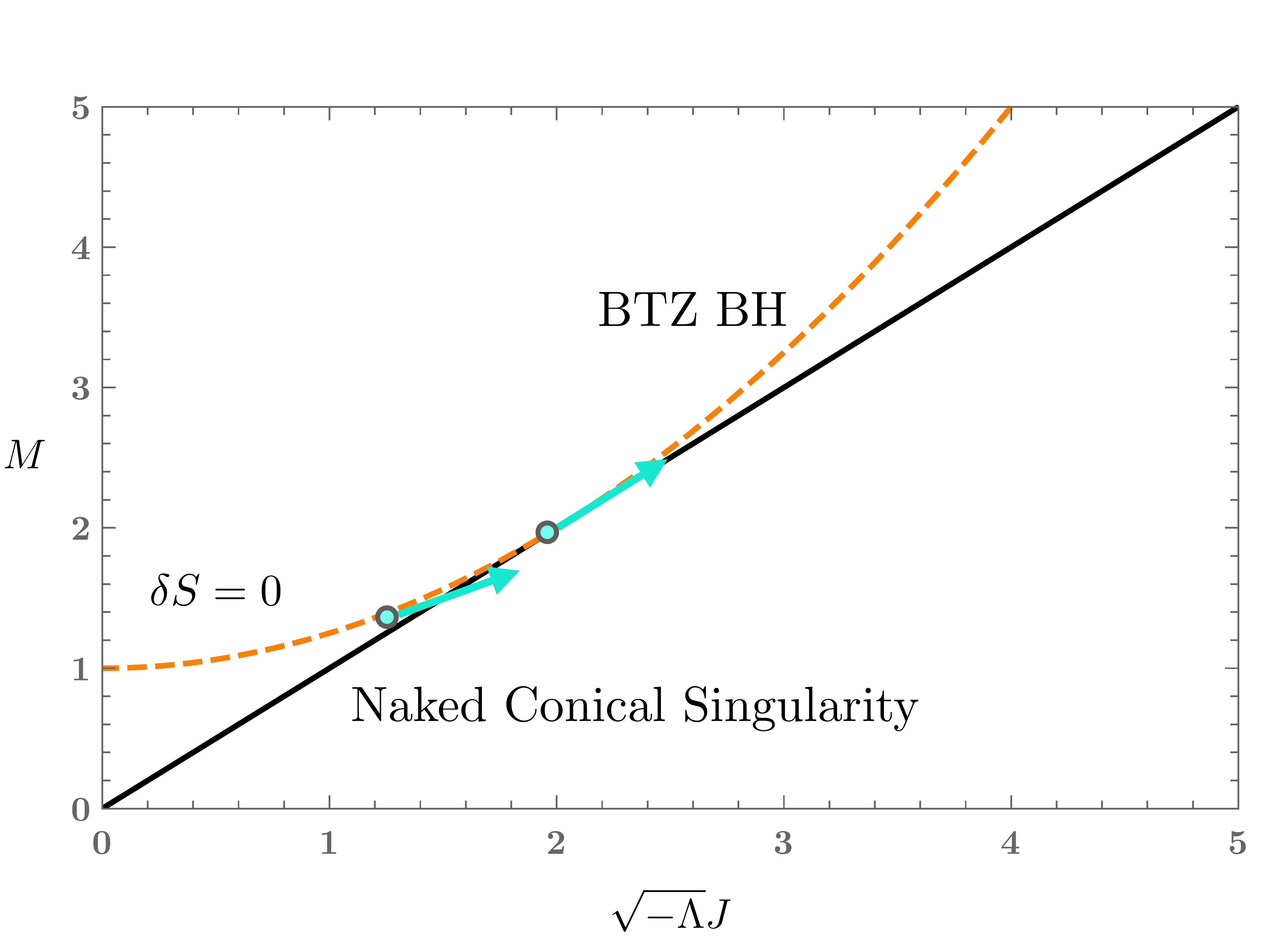}
  \caption{The parameter space of BTZ solutions in the 3-dimensional Einstein gravity.  The black solid line corresponds to extremal BTZ black holes.  Any point above this line corresponds to a non-extremal BTZ black hole, while any point below the line is a naked conical singularity.  The orange dashed curve is one of the curves of constant entropy, which meets the extremal BTZ line tangentially.  The tangent vector at the point of an extremal BTZ black hole will always bring it to another extremal BTZ black hole.  However, starting from a slightly non-extremal BTZ black hole, to linear order, the tangent vector can perturb the spacetime to become a naked conical singularity.}
  \label{fig:hubeny-BTZ-Einstein}
\end{figure}

As shown in Fig.~\ref{fig:hubeny-BTZ-Einstein}, if the initial spacetime is an extremal BTZ black hole, a tangent vector at this point is also tangent to the line representing extremal BTZ solutions.  Therefore given extremality, the best one can do is to deform the black hole to another extremal BTZ black hole.  WCCC is then preserved and no naked singularities will form.  However, if one starts at a slightly non-extremal BTZ black hole, the tangent to the curve of constant entropy is possible to move the original point to another point located in the region representing naked conical singularities. This type of violation of WCCC is exactly the Hubeny-type violation, which can be found at the linear order for near-extremal black holes.  As we will see in the following discussions, a conclusive answer to whether this type of perturbations truly leads to a violation of WCCC requires calculations to the second order.

Now we consider a 1-parameter family of solutions $\f(\l)$, $\f_0=\f(0)$ is a nearly extremal BTZ black hole in 3-dimensional Einstein gravity.  We then expand $f(\l)$ in e.q.~\eqref{eq:f-func} to second order in $\l$,
\begin{align}
\label{eq:f-sec-order}
 f(\l) = &\lrbrk{ M^2 - \a^2 J^2 } + 
 2\l \lrbrk{ M\d M - \a^2 J \d J}  +  \\ \nn
  &+ \l^2\lrsbrk{ (\d M)^2 -\a^2(\d J)^2  +  
 M\d^2 M - \a^2 J\d^2J } +\cO(\l^3)\,,
\end{align}
where we have introduced a parameter $\a=\sqrt{-\L}$.
For convenience we also introduce a parameter $\e$ according to
\be
\e = \frac{r^2_+ - r^2_{+,\mathrm{extremal}}}{r^2_{+,\mathrm{extremal}}} = \frac{\sqrt{M^2 - \a^2 J^2}}{M}\,.
\ee
The background spacetime corresponds to $\e \ll 1$, and $\e \rightarrow 0$ is the extremal limit.  
The null energy condition for the matter fields yields $\d M - \Omega_{\mathrm H} \d J\geq 0$, which is equivalent to the statement that black hole entropy always increases.  If we only consider perturbations to first order in $\l$, that entropy always increases will constrain $f(\l)$ by
\be
\label{eq: violation-einstein}
f(\l) \geq  M^2 \e^2 - 
 2\l\e \lrbrk{ \a^2 J\d J} +\cO(\l^2) \,.
\ee
It is then evident from this inequality that, when $\d J \sim \e M/\a$, it is possible to make $f(\l)<0$ by some careful choice of $\d J$.  This is exactly the Hubeny-type violation of WCCC.  The problem is that when $\d J \sim \e M/\a$, the violation of $f(\l)\geq 0$ is of order $M^2\e^2 \sim \a^2 (\d J)^2$, which is not fully captured to first order in $\l$.  Therefore to determine whether there is a true violation of WCCC, one needs to calculate all quantities in e.q.~\eqref{eq: violation-einstein} to the appropriate order.

We now consider the second order variations in order to give a bound for $f(\l)$.  Given the null energy conditions for the falling matter, we can obtain the following relation from the second order variational identity with no inner boundary contributions~\eqref{2ndVarId},
\be \label{eq:2ndVarEin}
\d^2 M - \Om_{\rm H} \d^2 J \geq \cE_{\S}(\f;\delta \f)\,,
\ee
where the canonical energy $\cE_\S$ is given by
\begin{align} \label{eq:canon-erg}
\cE_{\S}(\f;\d \f) 
&= \cE_{\cH}(\f;\d \f)+\cE_{\S_1}(\f;\d \f) \\ \nonumber
&=\int_{\cH} \Om(\f,\d \f, \mathscr{L}_\xi \d \f) 
  + \int_{\S_1} \Om(\f,\d \f, \mathscr{L}_\xi \d \f)\,.
\end{align}
In (3+1)-dimension, the term $\cE_{\cH}(\f;\d \f)$ is identified as the total flux of gravitational wave energy into the black hole~\cite{Hollands:2012sf}.  In (2+1)-dimensional Einstein gravity, however, there is no propagating degree of freedom in the bulk, i.e. there is no gravitational wave solution. Thus $\cE_{\cH}(\f;\d \f)=0$.  A more rigorous way to see this can be done by following the calculation of the canonical energy as in~\cite{Hollands:2012sf}, and we similarly find that
\be
\int_{\cH} \Om(\f,\d \f, \mathscr{L}_\xi \d \f) = \frac{1}{4\pi} \int_\cH (\kappa u) \d \s_{ab} \d \s^{ab} \hat{\e} + \frac{1}{16\pi} \int_{\cH\cap\S_1} (\kappa u )\d g^{ab} \d \s_{ab} \hat{\e}\,,
\ee
where $\k$ is the surface gravity, $u$ is an affine parameter on the future horizon, $\d \s_{ab}$ is the perturbed shear of the horizon generators, and $\hat{\e}$ is the volume element. In three dimension, it is found that every null geodesic congruence is shear-free~\cite{Chow:2009vt}, i.e. $\s_{ab}=0$, therefore $\d \s_{ab}=0$ on $\cH$ and the canonical energy on $\cH$ vanishes.  

Then we only need to calculate the canonical energy on $\S_1$.  According to our assumption, the perturbation $\d \f$, as induced by the falling matter, approaches a perturbation $\d \f^{\rm BTZ}$ towards another BTZ  black hole on $\S_1$.  Also since $\d \f^{\rm BTZ}$ has no gravitational wave energy through $\cH$, we may replace $\S_1$ by $\S$ and obtain that
\be
\cE_{\S_1}(\phi;\delta \phi) = \cE_{\S}(\f;\d \f^{\rm BTZ})\,.
\ee
We use the general second order variational identity~\eqref{2nd1} on this $\S$.  As before, we consider a one-parameter family of BTZ black holes, $\f^{\rm BTZ}(\b)$.  The black hole mass and angular momentum are given by $M(\b) = M+\b \d M^{\rm BTZ}$ and $J(\b) = J+\b \d J^{\rm BTZ}$, where $\d M^{\rm BTZ}$ and $\d J^{\rm BTZ}$ are fixed by the first order perturbation
for $\f(\l)$.  Therefore for this family of solutions, we have $\d^2 M=\d^2J=\d E=\d^2 C=0$.  In Eq.~\eqref{2nd1}, the only nonvanishing contribution in the evaluation of the canonical energy $\cE_{\S}(\f;\d \f^{\rm BTZ})$ then comes from the integral over the bifurcation surface $B$, which yields
\be
 \cE_{\S_1}(\phi;\delta \phi) = - T_{\rm H} \d^2 S^{\rm BTZ}\,.
\ee
Here, the minus sign is due to the fact that the bifurcation surface is the inner boundary of $\Sigma$. 

With the canonical energy being calculated,~\eqref{eq:2ndVarEin} now reads
\be
\label{eq:sec-var-ineq}
\d^2 M - \Om_{\rm H} \d^2 J \geq -T_{\rm H} \d^2 S^{\rm BTZ} \,.
\ee
Here the temperature of the BTZ black hole is given by
\be
T_{\rm H} = -\frac{\L (r^2_+ -r^2_-)}{2\pi r_+} = \frac{\a M \e}{\pi\sqrt{2M(1+\e)}}\,.
\ee
The second order variation of the black hole entropy is calculated as
\begin{align}
\d^2 S^{\rm BTZ} & = (\d J)^2\lrbrk{-\frac{\pi  \a   M \lrsbrk{\a ^2 J^2 (3 \e +2)+2 M^2 \e ^2 (\e +1)}}{\sqrt{2} \e ^3 \lrsbrk{M^3 (\e +1)}^{3/2}} } \\ \nn
&+(\d J \d M )\lrbrk{ \frac{\pi  \sqrt{2} \a  J (\e +2)}{M \e ^3 \sqrt{M^3 (\e +1)}} }
+(\d M )^2 \lrbrk{\frac{\pi (\e -2) (\e +1)}{\sqrt{2} \a  \e ^3 \sqrt{M^3 (\e +1)}} } \,.
\end{align}
where we have used the relation that for this family of solutions, $\d^2 M=\d^2J=0$. We assume that the first order perturbation is optimally done, i.e. $\d M = \Om_{\rm H} \d J$, and we use the inequality~\eqref{eq:sec-var-ineq} to constrain $f(\l)$ in e.q.~\eqref{eq:f-sec-order}.  We obtain that
\be
f(\l) \geq  M^2 \e^2 - 
 2\l\e\lrbrk{ \a^2 J\d J} 
 + \l^2\frac{\a^4J^2 (\d J)^2}{M^2}+
 \cO(\l^3,\e\l^2,\e^2\l, \e^3) \,,
\ee
which can be further written as
\be\label{thirdlaw}
f(\l) \geq \lrbrk{M\e - \l \frac{\a^2 J \d J}{M}}^2+
 \cO(\l^3,\e\l^2,\e^2\l, \e^3) \,.
\ee
Consequently, $f(\l)\geq 0$ when second order variations in $\l$ are also taken into account. Our gedanken experiment cannot destroy a near-extremal BTZ black hole in three-dimensional Einstein gravity, thus WCCC is preserved.

%%%%%%%%%%%%%%%%%    5. Conclusion    %%%%%%%%%%%%%%%%%%%%

\section{Conclusions and Discussions}
\label{sec:5}

  Along the line of Wald's proposals  \cite{Wald:I,Sorce:2017dst} for 4D Einstein gravity, in this paper we have considered the gedanken experiments of destroying a BTZ black hole for three different limits of Mielke-Baekler (MB) model of 3D gravity. They are (i) Einstein gravity, (ii) chiral gravity and (iii) torsional chiral gravity.  All three limits are free of perturbative ghosts and show different behaviors in the gedanken experiments.  We find that there are Hubeny-type violations for Einstein gravity but none for chiral gravity when trying to destroy a nonextremal BTZ black hole.  However, in these two theories, the WCCC holds for both extremal and nonextremal BTZ black holes if the falling matter obeys the null energy condition. It is philosophically interesting to see that WCCC prevails here even though the BTZ singularity is just a conical one. 
  
   On the other hand, for the torsional chiral gravity there is an additional contribution to the null energy condition from the spin angular momentum tensor even at the linear order of variations. Thus,  that the WCCC will hold or not depends on the imposition of additional null energy-like condition for the spin angular momentum tensor. If WCCC does not hold for the first order variations, one needs to check the  second order variation to see if there is Hubeny-type violation. However, the full formalism of deriving the second order variational identity for MB model is out of scope of this paper, and it deserves as a future work. 
  
  The third law of black hole dynamics was first proposed by Israel and a sketchy proof was also given \cite{Israel:1986gqz}, which states that one cannot turn a nonextremal black hole into an extremal one by throwing the matter in a finite time interval.  Later, the detailed proof was given by Sorce and Wald \cite{Sorce:2017dst} as described and adopted in this paper. In the context of AdS/CFT correspondence, the temperature of the boundary CFT is the same as the Hawking temperature of the black hole in the  bulk. Thus, our results in this paper can serve as an operational proof of thermodynamic third law by holographically mapping our gedanken experiments around a near-extremal BTZ black hole into the cooling processes of the boundary CFT toward zero temperature. Our generalization to BTZ black holes though seems straightforward, its implication to the third law of thermodynamics for holographic condensed matter systems is nontrivial and deserves further study. Especially, the generalization to the higher dimensional AdS black holes for more general gravity theories will give holographic tests of the third law of thermodynamics for the more realistic systems. We plan to attack this problem in the near future.
  
  Before ending the paper, we comment on one more point about the proof of the third law by noticing that the equality of (\ref{thirdlaw}) holds for one particular choice of parameter $\lambda$. This implies that one can reach the extremal black hole at the second order for this particular case. To pin down the issue, one needs to check the third order of variation for this particular $\lambda$ value. This is  too involved to carry out just for a measure-zero possibility. However, it is still an interesting issue for the future work.

%%%%%%%%%%%%%%%%%    Acknowledgements    %%%%%%%%%%%%%%%%%%%%

\subsection*{Acknowledgements}
BC acknowledges the support from the Brinson Foundation, the Simons Foundation (Award Number 568762), and the National Science Foundation, Grants PHY-1708212 and PHY-1708213. FLL is supported by Taiwan Ministry of Science and Technology through Grant No.~103-2112-M-003-001-MY3. BN is  supported by the National Natural Science Foundation of China with Grant No.~11505119.

%%%%%%%%%%%%%%%%%        Refer      %%%%%%%%%%%%%%%%%%%%

\providecommand{\href}[2]{#2}\begingroup\raggedright\endgroup

\end{document}